# Stress - strain rate relation in plug-free flow of dense granular fluids – a first-principles derivation

Moshe Schwartz[1,3] and Raphael Blumenfeld[2,3]

1. Beverly and Raymond Sackler School of Physics and Astronomy, Tel Aviv University, Ramat Aviv 69934, Israel
2. Earth Sciences and Engineering, Imperial College, London, UK
3. Cavendish Laboratory, University of Cambridge, UK

Email: rbb11@cam.ac.uk

We derive the macroscopic stress tensor for plug-free dense granular flow, using a first-principles coarse-graining of the intergranular forces. The derivation is based on the assumption, which defines the da Vinci Fluid model, that the intergranular interactions are dominated by normal contact forces and solid friction. An explicit form for the stress -- strain rate relation in the slow flow regime is obtained, providing, together with previously derived equations for the formation and growth dynamics of plug regions, a full closure for the rheology of dense granular fluids, in terms of well-defined material parameters. This relation allows us to quantify the strain rate, at which the flow crosses over from solid-friction-dominated to viscosity-dominated flow.

The significance of flow of dense granular matter to many natural and man-made phenomena cannot be over-emphasised. Modelllng the rheology in this regime is an important problem in the field [1,2] and a progress on it, especially from first-principles, would form the basis for further dynamic theories. The reproducible macroscopic patterns of dense flow suggest that, as for conventional fluids, it should be possible to construct continuum flow equations to model this rheology. Traditional modelling of flow, based on the theory of dense gases, is useful only when the concept of inter-particle collisions is meaningful and they lose validity for dense granular flow, where particles are mostly in prolonged contacts. Yet, since even in dense flows the intergranular forces are well understood, it is plausible that continuum flow equations can be obtained under appropriate coarse-graining, including a stress tensor with the right transformation properties under rotations. The aim of this paper is to go beyond previous empirical and phenomenological formulations and derive from basic considerations the dependence of the stress tensor on the strain rate in such flows.

The stress – strain rate relation in such flows is determined by intergranular interactions. Those consist of normal forces, which do not dissipate energy and energy dissipating forces predominantly through solid friction, described first by da Vinci [3], and later by Amontons [4] and Coulomb [5]. This is in contrast to ordinary fluids, where the dissipation is by viscosity. One of the main differences between conventional and dense granular fluids is that the latter are prone to formation of plugs, i.e. regions that move as macroscopic rigid objects. Plug regions (PRs) play an important role in the rheology of granular fluids and have been the focus of two recent papers [6,7]. To complete the description in those references, we focus here on deriving the stress tensor in plug-free regions of dense granular flow.

The role of inter-particle solid friction, as a significant dissipation mechanism in dense granular flow, has been recognised since the 80s. In particular, much research focused on identifying the constitutive relation between the stress and the strain rate in such flows, using several different approaches. One was an empirical conjecture, consistent with solid friction as the only dissipation mechanism [8,9], supplemented by the assumption of incompressibility. The actual form of the stress tensor was only marginally relevant in those works. A different approach was taken by da Cruz et. al. [10], who used two-dimensional simulations of disks, dissipating energy via both solid friction and inelastic collisions. The general form of the stress – strain rate relation is similar to that suggested in [8,9]. However, in contrast to the assumption in the previous works, they found that the local effective friction coefficient, $\mu$, is non-constant, depending linearly on the inertial number, $I$, which in turn is proportional to the norm of the strain rate tensor and inversely proportional to the square root of the pressure. Jop et. al. [11] followed with extensive numerical simulations in certain geometries [1] and obtained a scalar stress - strain rate relation. They then extended this relation to a three-dimensional (3D) tensorial form, complemented it with the incompressibility assumption and checked the formalism against experiments in six different geometries. Although the general form of their proposed stress tensor was the same as in the previous works, its dependence on $I$ made the formalism richer than the one obtained in [10]. Later, Kamrin and Koval [12]. and Bouzid et. al. [13] introduced non-local terms into the stress tensor both to correct for certain experimentally known geometric effects and to account for results obtained in numerical simulations.

Those developments led to the question of ill-posedness of the flow equations of fluids supporting such stress tensors, in particular under high wave vector disturbances. It was found that incompressible such models are mathematically ill-posed [8,14], when the effective friction coefficient is constant [8.9], but that they can be regularised under some conditions when $\mu$ depends on $I$ [14,15]. It was also shown that such $\mu(I)$ rheology can be fully regularised by introducing compressibility in a particular way [15-17].

Here we derive the stress - strain rate relation from first-principles by coarse-graining from the grain-scale. We also argue that the solid-friction dissipation dominates the low rate flow and identify the crossover to viscosity-dominated flow as a condition on the strain rate gradient. While we do not address the ill-posedness of the full flow equations, we discuss this issue in the concluding section and propose that its origin is physical, rather than in the mere form of the closure relation. Specifically, we suggest that it reflects an inherent instability of the flow to formation of plugs, and that the full rheology of dense granular fluids must include both plug and plug-free regions [6,7].

To derive a large-scale first-principles relation between the local stress and strain rate, we need to coarse-grain the intergranular interactions into an interaction between adjacent volume elements of the fluid. Before getting down to this task, it is important to comment that any derivation of such a relation can be valid only in plug-free regions of the flow. This is because flows in systems, in which dissipation is dominated by solid friction, are unstable towards formation and growth of plug regions [6]. Thus, even a highly accurate form of the stress tensor for plug-free regions is ultimately incomplete for a full description of the rheology, which should include the equation of motion and growth of the solid-like plug regions [7].

Our approach is based on separating the contributions to the stress tensor from the normal contact forces between volume elements, $\boldsymbol{\sigma}^{(n)}$, and from friction forces, $\boldsymbol{\sigma}^{(f)}$, and deriving an expression relating $\boldsymbol{\sigma}^{(f)}$ to $\boldsymbol{\sigma}^{(n)}$ and the strain rate tensor $\mathbf{T}$.

Consider a dense system of roughly spherical rigid convex grains, of typical size $d$, interacting via normal and frictional contact forces. We focus first on the intergranular tangential friction forces. Let $k$ denote a pair of grains, $i$ and $j$,

$\mathbf{N}_k$ be the normal contact force that grain $i$ applies to grain $j$, and $\mathbf{r}_{ij}$ be the vector extending from the centre of grain $i$ to its contact point with grain $j$. The friction force that $i$ applies to $j$ depends on the relative velocities of grains $i$ and $j$ at the contact point,

$$\mathbf{\Delta}_k = \mathbf{\Lambda}_k + \frac{1}{2}(\boldsymbol{\omega}_i + \boldsymbol{\omega}_j) \times \mathbf{R}_k + \frac{1}{2}(\boldsymbol{\omega}_i - \boldsymbol{\omega}_j) \times \boldsymbol{\rho}_k, \tag{1}$$

where $\mathbf{\Lambda}_k = \mathbf{v}_i - \mathbf{v}_j$ is the grains' relative centre of mass velocity, $\boldsymbol{\omega}_i$ and $\boldsymbol{\omega}_j$ are the angular velocities of the two grains around their respective centres of mass, $\mathbf{R}_k = \mathbf{r}_{ij} - \mathbf{r}_{ji}$ is the vector from the centre of mass of $i$ to that of $j$ and $\boldsymbol{\rho}_k = \mathbf{r}_{ij} + \mathbf{r}_{ji}$. Note that $\boldsymbol{\rho}_k = 0$ when $|\mathbf{r}_{ij}| = |\mathbf{r}_{ji}|$, which is the case for identical spherical grains.

When the grains rub against one another, the friction force, applied by grain $i$ to grain $j$, is described by the da Vinci - Amontons - Coulomb law:

$$\mathbf{F}_k = \mu_d |\mathbf{N}_k| \hat{\mathbf{u}}_k, \tag{2}$$

where $\mu_d$ is the dynamic friction coefficient between the members of the pair, and $\hat{\mathbf{u}}_k = \mathbf{\Delta}_k / |\mathbf{\Delta}_k|$ is a unit vector. For simplicity, we assume the same $\mu_d$ between all rubbing particles. When $\mathbf{\Delta}_k = 0$, we only know that the inter-granular friction force satisfies $\mathbf{F}_k \cdot \mathbf{N}_k = 0$ and $|\mathbf{F}_k| \leq \mu_s |\mathbf{N}_k|$, where $\mu_s (\geq \mu_d)$ is the static friction coefficient. We shall see below that $\mu_s$ does not play any role in the plug-free stress tensor.

Our aim is to obtain, by coarse-graining, the effective interaction between two adjacent volume elements of the fluid. The volume elements are regarded as sufficiently large to contain many grains, but to be much smaller than the system size. Consider two such volume elements, $V_A$ and $V_B$ separated by an imaginary plane. The plane may cut individual grains, which are then deemed to belong to either $V_A$ or $V_B$, depending on the locations of their centres of mass (see figure 1).

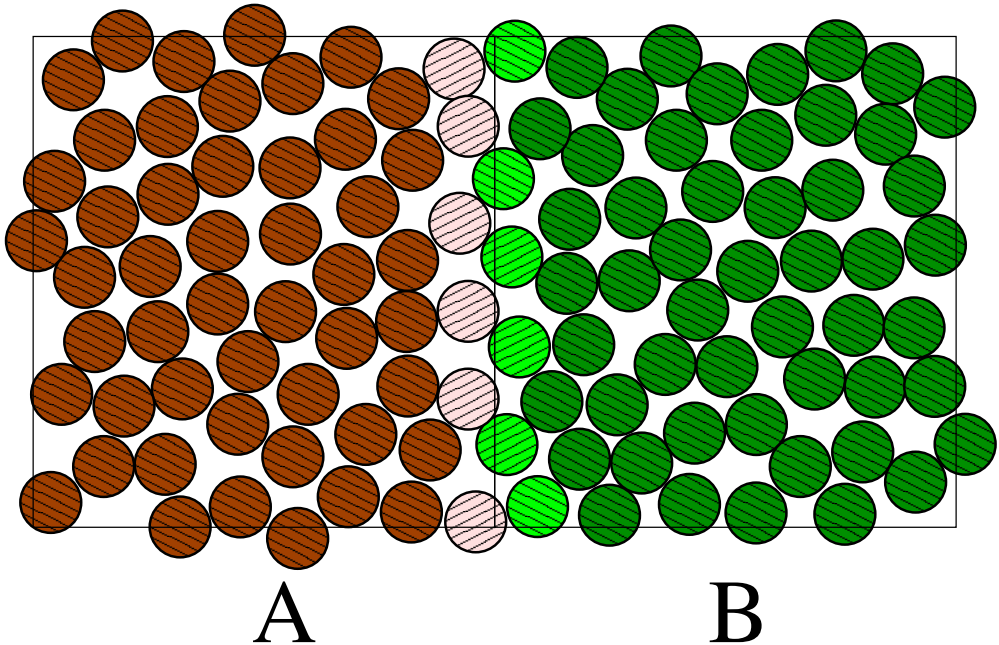

Figure 1: The boundary grains (lighter shades) to the left of the `imaginary' plane between volume elements $A$ and $B$ are in contact with, and apply forces to, grains to its right.

The first step is to obtain an expression for the net force per unit area applied by grains in $V_A$ to grains in $V_B$ around a point $\mathbf{x}$ on the boundary plane. We consider separately the contributions to this force from the non-dissipative normal contact forces and from the dissipative frictional (tangential) forces, between pairs across the boundary plane. By normal and tangential we refer here to directions with respect to the individual inter-granular tangent contact planes. The average normal force per unit area applied by $V_A$ to $V_B$ is

$$\mathbf{v}(\mathbf{x}) = \pi(\mathbf{x})\langle\mathbf{N}\rangle \equiv \pi(\mathbf{x})\mathbf{N}(\mathbf{x}), \qquad (3)$$

where $\langle\mathbf{N}\rangle$ is the spatial average of the normal forces applied by the $A$ members of the grain pairs to their $B$ members and $\pi(\mathbf{x})$ is the number density of pairs per unit area at $\mathbf{x}$. This average is over a circular area, around the point $\mathbf{x}$, which is large enough to contain a statistically significant large number of such pairs, but can be considered macroscopically small. Similarly, the average solid friction force per unit area, which volume element $A$ applies to $B$, is $\boldsymbol{\varphi}(\mathbf{x}) = \pi(\mathbf{x})\mathbf{F}(\mathbf{x})$, where $\mathbf{F}(\mathbf{x}) = \langle\mathbf{F}\rangle$ is the average friction force the $A$ members of the pairs applies to the $B$ members in the vicinity of $\mathbf{x}$.

To obtain the two contributions to the total stress tensor, $\boldsymbol{\sigma}^{(n)}(\mathbf{x})$ and $\boldsymbol{\sigma}^{(f)}(\mathbf{x})$, while ensuring their proper tensorial nature, we must introduce two more planes, orthogonal to the original plane and to one another. Denoting this triad of planes by $\alpha = 1, 2, 3$ and the Cartesian components of $\mathbf{v}_\alpha(\mathbf{x})$ and $\boldsymbol{\varphi}_\alpha(\mathbf{x})$, where each of the components is perpendicular to one of the planes by $\beta = 1, 2, 3$, it is straightforward to see that

$$\boldsymbol{\sigma}^{(n)}_{\alpha\beta}(\mathbf{x}) = \mathbf{v}_{\alpha\beta}(\mathbf{x}) \text{ and } \boldsymbol{\sigma}^{(f)}_{\alpha\beta}(\mathbf{x}) = \boldsymbol{\varphi}_{\alpha\beta}(\mathbf{x}). \qquad (4)$$

Next, we consider one of the planes and evaluate $\mathbf{F}(\mathbf{x})$ by taking the spatial average of $\mu_d |\mathbf{N}_k| \hat{\mathbf{u}}_k$. Noting that there is no correlation between the magnitude of the normal force and the direction of the velocity difference between the members of the pair across the plane, we have

$$\mathbf{F}(\mathbf{x}) = \mu_d \langle |\mathbf{N}|\hat{\mathbf{u}}\rangle = \mu_d \langle |\mathbf{N}|\rangle\langle\hat{\mathbf{u}}\rangle. \qquad (5)$$

To calculate $\langle\hat{\mathbf{u}}\rangle$, we separate $\boldsymbol{\Lambda}_k$ into its average $\langle\boldsymbol{\Lambda}_k\rangle$, and a fluctuation,

$\delta \boldsymbol{\Delta}_k d$. The reason for including the typical grain size $d$, in the definition of the fluctuations will become clear below. The average of the difference of center of mass velocities Is

$$\langle \boldsymbol{\Lambda}_k \rangle = -\langle \hat{\mathbf{e}}_k \rangle \cdot \nabla \mathbf{v}(\mathbf{x}) d . \tag{6}$$

The average of the second term on the right hand side of eq. (1) is

$$\left\langle \frac{1}{2}(\boldsymbol{\omega}_i + \boldsymbol{\omega}_j) \times \mathbf{R}_k \right\rangle = -\langle \hat{\mathbf{e}}_k \rangle \times \boldsymbol{\omega}(\mathbf{x}) d , \tag{7}$$

where $\hat{\mathbf{e}}_k = \mathbf{R}_k / |\mathbf{R}_k|$ is a unit vector pointing from the center of mass of grain $i$ to the center of mass of grain $j$, $\mathbf{v}(\mathbf{x})$ is the coarse-grained velocity field and $\boldsymbol{\omega}(\mathbf{x}) = \nabla \times \mathbf{v}(\mathbf{x})$ is the local macroscopic angular velocity of the fluid. We make the (plausible) assumption that the third term on the right hand side of eq. (1) averages to zero. Combining (6) and (7), it is readily verified that,

$$\langle \boldsymbol{\Delta}_k \rangle = -\langle \hat{\mathbf{e}}_k \rangle \cdot \mathbf{T} d \equiv \boldsymbol{\Delta} d . \tag{8}$$

The right hand side of eq. (8) contains, in addition to the strain rate tensor, $\mathbf{T}$, the average of $\hat{\mathbf{e}}_k$. The latter does not vanish, because $\hat{\mathbf{e}}_k$ has always a component pointing from the center of the volume element $V_A$ to that of $V_B$.

To evaluate $\langle \hat{\mathbf{u}} \rangle$, we use the identity $\nabla_\mathbf{w} |\mathbf{w}| \equiv \mathbf{w}/|\mathbf{w}|$, where $\mathbf{w}$ is an arbitrary vector, to write

$$\langle \hat{\mathbf{u}} \rangle = \nabla_{\boldsymbol{\Delta}(\mathbf{x})} \left\langle \left| \boldsymbol{\Delta}(\mathbf{x}) + \delta \boldsymbol{\Delta}_k \right| \right\rangle , \tag{9}$$

where $\delta \boldsymbol{\Delta}_k \equiv \delta \boldsymbol{\Delta}_k d$. Next, we assume that the average $\langle |\boldsymbol{\Delta}_k| \rangle$ does not depend on the specific choice of the separating plane and that it can be expressed in a rotational covariant form:

$$\left\langle \left|\mathbf{\Delta}_k\right|\right\rangle = \left\langle \{\frac{1}{3}[\sum_{\alpha=1}^{3} \mathbf{\Delta}_{k\alpha}]^2\}^{1/2}\right\rangle \equiv \left\langle \left|\mathbf{\Delta}_{\bar{k}}\right|\right\rangle . \tag{10}$$

This form involves all the three orthogonal planes discussed above, where $k\alpha$ denotes a pair of grains traversing the plane $\alpha$ and $\bar{k}$ is a shorthand notation for the triad $\{k1, k2, k3\}$. We emphasize that the average is taken now over all pairs traversing the areas of the three perpendicular circles passing through $\mathbf{x}$.

Defining the norms $\left|\mathbf{\Delta}(\mathbf{x})\right| = [\sum_{\alpha,\beta=1}^{3} \mathbf{\Delta}_{\alpha\beta}^2(\mathbf{x})]^{1/2}$, $\left|\delta\mathbf{\Delta}_{\bar{k}}\right| = \{\frac{1}{3}[\sum_{\alpha=1}^{3} \delta\mathbf{\Delta}_{k\alpha}]^2\}^{1/2}$

and averaging, we obtain

$$\left\langle \left|\mathbf{\Delta}_{\bar{k}}\right|\right\rangle = \frac{1}{\sqrt{3}} f(\psi) \left|\mathbf{\Delta}(\mathbf{x})\right| , \tag{11}$$

where

$$\psi \equiv \left|\mathbf{\Delta}(\mathbf{x})\right| / \left\langle \left|\delta\mathbf{\Delta}_{\bar{k}}\right|\right\rangle . \tag{12}$$

Using relations (9), (11) and (12) for all the planes, $\alpha = 1, 2, 3$, we have

$$\left\langle \hat{\mathbf{u}}_{\alpha\beta}\right\rangle = \frac{\partial}{\partial \mathbf{\Delta}_{\alpha\beta}(\mathbf{x})} \left\langle \left|\mathbf{\Delta}_k\right|\right\rangle = g(\psi) \frac{\mathbf{\Delta}_{\alpha\beta}(\mathbf{x})}{\left|\mathbf{\Delta}(\mathbf{x})\right|} . \tag{13}$$

Multiplying and dividing the right hand side of eq. (13) by $\left\langle \left|\mathbf{N}\right|\right\rangle$ and using relation (8) we obtain,

$$\left\langle \hat{\mathbf{u}}_{\alpha\beta}(\mathbf{x})\right\rangle = -g(\psi) \frac{\mathbf{N}_{\alpha\gamma}(\mathbf{x})\mathbf{T}_{\gamma\beta}(\mathbf{x})}{\left|\mathbf{N}_{\gamma\delta}(\mathbf{x})\mathbf{T}_{\delta\gamma}(\mathbf{x})\right|} . \tag{14}$$

From (11)-(13), it is straightforward to obtain $g(\psi) = \frac{1}{\sqrt{3}}[f(\psi) + \psi f'(\psi)]$.

Multiplying now both sides of (14) by $\mu_d \pi(\mathbf{x}) |\langle \mathbf{N}(\mathbf{x}) \rangle|$, yields the required expression for the solid friction contribution to the stress tensor (see eq. (4)), expressed in terms of the normal contact stress tensor and the strain rate tensor,

$$\sigma_{\alpha\beta}^{(f)}(\mathbf{x}) = -\mu_d \phi g(\psi) |\sigma^{(n)}(\mathbf{x})| \frac{\sigma_{\alpha\gamma}^{(n)}(\mathbf{x}) T_{\gamma\beta}(\mathbf{x})}{|\sigma^{(n)}(\mathbf{x}) T(\mathbf{x})|} , \qquad (15)$$

where $\phi \equiv \langle |\mathbf{N}_{\bar{k}}| \rangle / |\mathbf{N}(\mathbf{x})|$ is expected to depend only on the local density.

To illustrate the usefulness of this derivation, we apply it to the simple case of an idealized incompressible granular flow, a case discussed previously in the literature [8,11]. In this case, $\phi$ is constant throughout the system and we simplify the notation: $\bar{\mu} \equiv \mu_d \phi$. For clarity, we also assume that $\bar{\mu} \ll 1$ and derive the total stress tensor to first order in $\bar{\mu}$. To obtain $\boldsymbol{\sigma}^{(f)}(\mathbf{x})$ to this order, it is sufficient to consider the zeroth order of $\boldsymbol{\sigma}^{(n)}(\mathbf{x})$ on the right-hand side of eq. (15), $\boldsymbol{\sigma}_0^{(n)}(\mathbf{x})$. But the latter is nothing but the stress tensor of an ordinary, incompressible fluid:

$$\boldsymbol{\sigma}_0^{(n)}(\mathbf{x}) = -p_0(\mathbf{x}) \mathbf{I} + \eta \mathbf{T}(\mathbf{x}) , \qquad (16)$$

where $p_0$ is the pressure of the ordinary fluid, $\eta$ its viscosity and $\mathbf{I}$ the unit tensor.

In the following, we apply the above analysis to flow under low shear rate. Significantly, eq. (15) shows that, in such flows, the solid friction contribution to the stress tensor is of zero order in the strain rate. Therefore, solid friction dominates over the viscous contribution, which is first order in the strain rate. To first order in $\bar{\mu}$, the total stress tensor, $\sigma^{(t)}(\mathbf{x}) = \sigma^{(n)}(\mathbf{x}) + \sigma^{(f)}(\mathbf{x})$, is then

$$\boldsymbol{\sigma}^{(t)} = -(P_0 + \bar{\mu} P_1)\mathbf{I} + \bar{\mu} g(\psi) P_0 \frac{\mathbf{T}}{|\mathbf{T}|} + \eta \nabla \cdot \mathbf{T} . \qquad (17)$$

In this expression, the first order correction to the pressure, $P_1$, is determined from the incompressibility condition, which results in the following equation:

$$-\nabla P_1 + [\nabla \cdot g(\psi) P_0 \frac{\mathbf{T}}{|\mathbf{T}|}]_L = 0 , \qquad (18)$$

where the subscript $L$ denotes the longitudinal part of a vector field. We do not dwell on eq. (18) because, although the pressure depends on the velocity field, the latter is not affected by the first order correction in $\bar{\mu}$ to the pressure, as evident from eq. (17). This is no different than in ordinary incompressible liquids, where the velocity field determines the pressure, up to a constant, but the pressure does not affect the velocity field at all.

We included in eq. (18) the viscosity contribution to the stress for comparison purposes. This contribution is overwhelmed by the solid friction one at low strain rates, but it dominates at high strain rate and its gradients. The two are comparable when

$$|\nabla \cdot \mathbf{T}| \approx \frac{\bar{\mu} g(\psi) P_0}{\eta} . \qquad (19)$$

As we show below, $g(\psi)$ can be an estimated at low strain rates and, therefore, relation (19) quantifies the crossover between the two regimes. The viscosity term, which is negligible at low strain rates, may regularise the otherwise ill-posed equations. To see how, note that the ill-posedness implies unbounded growth of high-momentum perturbations, which leads to growth of the strain rate, and it is exactly such high momenta that the viscosity term suppresses.

Thus, in the low strain rate approximation, the last term on the right of eq. (17) may be neglected. Furthermore, it can be readily verified that, under the same approximations, the pressure $P_0$, which appears in the frictional term, can be replaced by $\bar{P}_0$, the average pressure in the system.

Next, we would like to express $\psi$, defined in eq. (12), in terms of the measurable strain rate tensor. Inspecting relation (17), we expect that, to zero order in $\bar{\mu}$, $\langle \hat{\mathbf{e}}_{\alpha\beta} \rangle = \varepsilon \delta_{\alpha\beta}$, where $0 < \varepsilon < 1$ is a dimensionless parameter that may depend (possibly weakly) on the density. It follows from relation (8) that $\psi = \varepsilon |\mathbf{T}| / \langle |\delta \mathbf{\Lambda}_{\bar{k}}| \rangle$. We assume that the internal fluctuations are only driven externally, e.g. by shear, and that $\psi$ tends to a positive constant $\psi_0$, as $|\mathbf{T}|$ tends to zero. Then, at very low macroscopic shear, we can replace $g(\psi)$, which may depend generally on the local density and the strain rate, by a constant $g(\psi_0)$. This simplifies eq. (17) and reduces it to the form conjectured by Schaeffer [8].

To extend the analysis beyond very low strain rates, we do the following. First, we need to obtain the form of $g(\psi)$ for small and large $\psi$. Then, from the definitions of $|\delta \mathbf{\Lambda}_{\bar{k}}|$ and $\psi$, we have that $f(\psi) = \psi^{-1}[1 + a\psi^2]$ for small $\psi$ and $f(\psi) = 1 + a'\psi^{-2}$ for large $\psi$. Here $a$ and $a'$ are positive dimensionless constants that depend on the specific distribution of $|\delta \mathbf{\Lambda}_{\bar{k}}|$. These relations follow from the generic form of the probability density function of $|\delta \mathbf{\Lambda}_{\bar{k}}|$, $\frac{P(|\delta \Delta_{\bar{k}}|/\langle|\delta \Delta_{\bar{k}}|\rangle)}{\langle|\delta \Delta_{\bar{k}}|\rangle}$. Using then the relation between $f$ and $g$, given below eq. (14), we have

$$g(\psi) = \begin{cases} \dfrac{2}{\sqrt{3}} a\psi & \text{for small } \psi \\ \dfrac{1}{\sqrt{3}}\left(1 - \dfrac{a'}{\psi^2}\right) & \text{for large } \psi \end{cases} \qquad (20)$$

While measuring $\psi$ experimentally is not easy, it is expected to depend only on a dimensionless scalar, which can be constructed from the strain rate tensor. A natural choice is the inertial number [10,11],

$$I = \frac{|\mathbf{T}|d}{\sqrt{p/\rho}}, \qquad (21)$$

where $\rho$ is the mass density of the grains. Using the inertial number has the advantage that it depends on the strain rate tensor, the pressure and readily measurable grain parameters. Thus, the form expected for $g(\psi)$ in eqs. (15) and (17) is

$$g(\psi) = c_0 + c(I), \qquad (22)$$

with $c(0) = 0$ and $c(I)$ a monotonically increasing function of $I$ that approaches asymptotically $1/\sqrt{3}$, as follows from eq. (20). The pre-factor on the right-hand side of eq. (15), $\mu(I) = \bar{\mu}[c_0 + c(I)]$, can be readily measured in setups, where the strain rate tensor is constant throughout the system, an example of which is reported in [11]. The value of $\mu(I)$ then serves as input to determine the total stress tensor in eqs. (15) and (17). This is no different than measuring the viscosity in ordinary fluids and using it as input to determine the stress tensor.

To conclude, we have derived, from first principles, the stress tensor of plug-free flow of dense granular fluids in the low strain rate regime, which is significant to many natural processes, technological applications and research disciplines. In this regime, the viscosity contribution to the stress is linear in the strain rate and, therefore, negligible compared to that of solid friction, which is a homogeneous function of degree zero in the strain rate. A novel result is the explicit dependence of the stress tensor on the grain-scale interaction statistics through the parameter $\psi$, relating persistent and random local behaviour. This makes the derived stress – strain rate relation an improvement on existing empirical and phenomenological proposals in the literature [8,11].

We also quantified the crossover in the nature of the flow from solid friction- to viscosity-dominated, which translates to a condition on the magnitude of the norm of the strain rate gradient, expressed in terms of more basic parameters.

We emphasise that our derivation is from first-principles and, as such, gives the correct stress - strain rate constitutive relation in the plug-free slow flow regime. However, it is essential to note that this regime is inherently unstable to formation and growth of plugs, as has been shown in [6,7]. This has two ramifications. One is that the problem of mathematical ill-posedness of equations of the form of (17) [8] may be resolved by combining the derivation here with the description of plug formation and dynamics [6,7]. This is anyway needed in order to construct a full theory of dense granular flow. The other ramification is that the strain rate vanishes inside plugs and, as these form and grow, the gradients of the strain rate at their boundaries increase. This means that there are three types of regions in such flows: plugs, wherein strain rate gradients vanish, plug-free regions, wherein solid friction dominates the rheology, and the boundary layers between the two, wherein the dissipation is viscosity-dominated. All these must also be taken into consideration in the full theory.

Further development of this model should include: (a) construction of numerical flow codes, incorporating this plug-free flow with plug formation and dynamics. (b) Improvements to the model by relaxing some of the approximations made here. One example is to consider higher friction coefficient. Another example is to extend the model to higher strain rates, when solid friction and viscous damping become comparable. Another possible extension is to determine the explicit dependence of $\psi$ on the inertial nummber, $I$. A furtther useful improvement to the model would be the relaxation of the assumption of the incompressibility. This assumption, used frequently in regular fluids, simplifies the analysis, but including effects of dilation and density-dependent friction coefficient are relevant to dense granular fluids. Such an extension could be related to our parameter $\phi$. All these are outside the scope of this paper. (c) Theoretical studies of flows, in various geometries and under a range of boundary conditions, that could be tested against real and numerical experiments. In particular, to determine the range fo validity of our results, it would also be useful to test numerically and experimentally where the transition occurs between the solid fiction- and viscosity-dominated types of flow. We are exploring some of these directions currently.

Acknowledgement:

MS acknowledges support the ISF within the ISF-UGC joint research program framework, grant number 839/14. RB acknowledges research funding support of National University of Defense Technology, Hunan, China.

**References**

[1]   Gdr MiDi, Eur. Phys. J.   **E 14**, 341 (2004).
[2]   Y. Forterre, O. Pouliquen, Annu. Rev. Fluid Mech.   **40**, 1 (2008).
[3]   L. da Vinci, *Static measurements of sliding and rolling friction*, Codex Arundel, folios 40v, 41r, British Library.
[4]   Amontons G., *Histoire de l'Academie Royale des Sciences avec les*

*Memoires de Mathematique et de Physique, 1699-1708*, (Chez Gerald Kuyper, Amsterdam, 1706-1709), p. 206.

[5] C. A. de Coulomb, *Theorie des machines simples, en ayant egard au frottement de leurs parties et A la roideur des cordages*, (reprinted by Bachelier, Paris 1821).

[6] R. Blumenfeld, S. F. Edwards and M. Schwartz, EuroPhys. J. **E 32**, 333 (2010).

[7] M. Schwartz and R. Blumenfeld, Granular Matter **13**, 241 (2011).

[8] D.G. Schaeffer, J. Diff. Eq. **66**, 19 (1987).

[9] G.I. Tardos, Powder Technology **92**, 61 (1997).

[10] F. da Cruz, S. Emam, M. Prochnow, J.-N. Roux and F. Chevoir, Phys. Rev. **E 72**, 021309 (2005).

[11] P. Jop, Y. Forterre and O. Pouliquen, Nature **441**, 727 (2006).

[12] K. Kamrin and G. Koval, Phys. Rev Lett. **108**, 178301 (2012).

[13] M. Bouzid, M. Trulsson, P. Claudin, E. Clément, and B. Andreotti, Phys. Rev. Lett. **111**, 238301 (2013).

[14] T. Barker, D. G. Schaeffer, P. Bohorquez and J. M. N. T. Gray, J. Fluid Mech. **779**, 794 (2015).

[15] T. Barker and J. M. N. T Gray, J. Fluid Mech. **828**, 5 (2017).

[16] T. Barker, D.G. Schaeffer, M. Shearer and J.M. N.T. Gray, Proc. R.Soc. A **473**: 20160846 (2017).

[17] J. Heyman, R. Delannay, H. Tabuteau and A. Valance, J. Fluid Mech. **830**, 553 (2017).